\documentclass[draft]{agujournal2019}
\usepackage{url} %this package should fix any errors with URLs in refs.
\usepackage{lineno}
\usepackage[inline]{trackchanges} %for better track changes. finalnew option will compile document with changes incorporated.
\usepackage{soul}
\usepackage[utf8]{inputenc}
\usepackage{csquotes}

\justifying
\draftfalse

\journalname{Geophysical Research Letters}

\begin{document}

%%%%%%%%%%%%%%%%%%%%%%%%%%%%%%%%%%%%%%%%%%%%%%%
%  TITLE
%
% (A title should be specific, informative, and brief. Use
% abbreviations only if they are defined in the abstract. Titles that
% start with general keywords then specific terms are optimized in
% searches)
%
%%%%%%%%%%%%%%%%%%%%%%%%%%%%%%%%%%%%%%%%%%%%%%%

% Example: \title{This is a test title}

\title{The forgotten role of wave dynamics in modulating the low cloud response to warm pool warming}
%%%%%%%%%%%%%%%%%%%%%%%%%%%%%%%%%%%%%%%%%%%%%%%
%
%  AUTHORS AND AFFILIATIONS
%
%%%%%%%%%%%%%%%%%%%%%%%%%%%%%%%%%%%%%%%%%%%%%%%

% Authors are individuals who have significantly contributed to the
% research and preparation of the article. Group authors are allowed, if
% each author in the group is separately identified in an appendix.)

% List authors by first name or initial followed by last name and
% separated by commas. Use \affil{} to number affiliations, and
% \thanks{} for author notes.
% Additional author notes should be indicated with \thanks{} (for
% example, for current addresses).

% Example: \authors{A. B. Author\affil{1}\thanks{Current address, Antartica}, B. C. Author\affil{2,3}, and D. E.
% Author\affil{3,4}\thanks{Also funded by Monsanto.}}

\authors{Cristian Proistosescu \affil{1,2}, Pappu Paul \affil{1}, Nicholas J. Lutsko \affil{3}, Andrew I.L. Williams\affil{3}, Malte F. Stuecker\affil{4,5}}

 \affiliation{1}{Department of Climate, Meteorology, and Atmospheric Sciences, University of Illinois Urbana Champaign, Urbana, IL}
 \affiliation{2}{Department of Earth Sciences and Environmental Change, University of Illinois Urbana Champaign, Urbana, IL}
 \affiliation{3}{Scripps Institution of Oceanography, La Jolla, CA
}
 \affiliation{4}{Department of Oceanography, School of Ocean and Earth Science and Technology (SOEST), University of Hawai`i at  M\={a}noa, Honolulu, HI}

\affiliation{5}{International Pacific Research Center (IPRC), SOEST, University of Hawai`i at  M\={a}noa, Honolulu, HI}

%\affiliation{1}{Urbana, Illinois}
%(repeat as many times as is necessary)

% Corresponding author mailing address and e-mail address:

% (include name and email addresses of the corresponding author.  More
% than one corresponding author is allowed in this LaTeX file and for
% publication; but only one corresponding author is allowed in our
% editorial system.)

% Example: \correspondingauthor{First and Last Name}{email@address.edu}

\correspondingauthor{Cristian Proistosescu}{cristi@illinois.edu}

%%%%%%%%%%%%%%%%%%%%%%%%%%%%%%%%%%%%%%%%%%%%%%%
% KEY POINTS
%%%%%%%%%%%%%%%%%%%%%%%%%%%%%%%%%%%%%%%%%%%%%%%
%  List up to three key points (at least one is required)
%  Key Points summarize the main points and conclusions of the article
%  Each must be 140 characters or fewer with no special characters or punctuation and must be complete sentences

% Example:
% \begin{keypoints}
% \item	List up to three key points (at least one is required)
% \item	Key Points summarize the main points and conclusions of the article
% \item	Each must be 140 characters or fewer with no special characters or punctuation and must be complete sentences
% \end{keypoints}

\begin{keypoints}
\item Convective Quasi-Equilibrium Weak Temperature Gradient theory does not explain the response of East Pacific low clouds to warm pool warming.

\item Warm pool warming leads to negative radiation anomalies in the tropics, but positive anomalies in the low cloud decks.

\item Green's Function experiments indicate planetary waves modulate the response of these low clouds.
%\item enter point 3 here
\end{keypoints}

%%%%%%%%%%%%%%%%%%%%%%%%%%%%%%%%%%%%%%%%%%%%%%%
%
%  ABSTRACT and PLAIN LANGUAGE SUMMARY
%
% A good Abstract will begin with a short description of the problem
% being addressed, briefly describe the new data or analyses, then
% briefly states the main conclusion(s) and how they are supported and
% uncertainties.

% The Plain Language Summary should be written for a broad audience,
% including journalists and the science-interested public, that will not have 
% a background in your field.
%
% A Plain Language Summary is required in GRL, JGR: Planets, JGR: Biogeosciences,
% JGR: Oceans, G-Cubed, Reviews of Geophysics, and JAMES.
% see http://sharingscience.agu.org/creating-plain-language-summary/)
%
%%%%%%%%%%%%%%%%%%%%%%%%%%%%%%%%%%%%%%%%%%%%%%%

%% \begin{abstract} starts the second page

\begin{abstract}
The Pattern Effect describes the dependence of top-of-atmosphere radiation anomalies on changes in the pattern of sea surface temperatures. The emerging consensus in the field explains the impact of Pacific warm pool temperature on radiation using Convective Quasi-Equilibrium Weak Temperature Gradient (QE-WTG) theory: warm pool warming leads to increase in free-tropospheric temperatures across the tropics, a strengthening of inversion, increased cloud cover in  the East Pacific low cloud decks, and negative radiative anomalies. Here we call on overlooked past results and new simulations from the Energy Exascale Earth System model to show that Rossby waves dominate the low-cloud response over the subtropical East Pacific low cloud decks, leading to decrease cloud cover in the low cloud decks. While the global radiative response is negative and consistent with QE-WTG, it is dominated by the response of the deep tropics, rather than the subtropical low cloud decks. 

\end{abstract}

\section*{Plain Language Summary}
The subtropical East Pacific is characterized by areas of dense low clouds that reflect a significant amount of sunlight, called the “low cloud decks”. These cloud decks are very important in regulating Earth’s energy balance and can amplify the amount of global warming caused by greenhouse gases. A topic of great interest in the recent literature is how these clouds respond to far away warming in tropical West Pacific. The field has coalesced around a paradigm that explains this response by invoking the fact that temperature gradients in the upper troposphere must be small. Here we show that paradigm leads to incorrect predictions and argue that we need to account for the presence of large-scale atmospheric motions caleld Rossby Waves. 
%ttps://www.agu.org/Share-and-Advocate/Share/Community/Plain-language-summary

%%%%%%%%%%%%%%%%%%%%%%%%%%%%%%%%%%%%%%%%%%%%%%%
%
%  BODY TEXT
%
%%%%%%%%%%%%%%%%%%%%%%%%%%%%%%%%%%%%%%%%%%%%%%%

%%% Suggested section heads:
% \section{Introduction}
%
% The main text should start with an introduction. Except for short
% manuscripts (such as comments and replies), the text should be divided
% into sections, each with its own heading.

% Headings should be sentence fragments and do not begin with a
% lowercase letter or number. Examples of good headings are:

% \section{Materials and Methods}
% Here is text on Materials and Methods.
%
% \subsection{A descriptive heading about methods}
% More about Methods.
%
% \section{Data} (Or section title might be a descriptive heading about data)
%
% \section{Results} (Or section title might be a descriptive heading about the
% results)
%
% \section{Conclusions}

%``For every complex problem there is an answer that is clear, simple, and wrong."
%\begin{flushright}
%H.L. Menken.
%\end{flushright}

\section{The Pattern Effect and the Convective Quasi-Equilibrium Weak Temperature Gradient approximation}

The Pattern Effect describes how changes in sea surface temperature (SST) patterns modulate the top-of-atmosphere (TOA) radiative budget and the net global radiative feedback \cite{rugenstein_patterns_2023}. SST patterns with relatively more warming in the tropical West Pacific lead to negative (i.e. outgoing) global-mean radiative anomalies and a more negative net feedback. This result is consistent when considering overall warming patterns in  climate models \cite{andrews_dependence_2015,andrews_effect_2022,zhou_impact_2016, fueglistaler_peculiar_2021, armour_sea-surface_2024} and observations \cite{fueglistaler_observational_2019,loeb_new_2020,myers_observational_2021}, as well as in Greens Function maps of the sensitivity of TOA radiation to localized warming \cite{zhou_analyzing_2017,dong_attributing_2019, zhang_using_2023, alessi_surface_2023, blochjohnson_greens_2024, wu_applying_2025, bloch-johnson_spatial_2020,kang_recent_2023,fredericks_estimating_2025, thompson_observational_2025}. There is also consensus that feedback changes are dominated by changes in low-cloud cover (LCC), themselves mediated by changes in the stability of the lower troposphere \cite{ceppi_refined_2019, tam_meteorological_2026}, usually quantified as Estimated Inversion Strength \cite<EIS,>[]{wood_relationship_2006}.

Particular attention has been given to the recent pattern of accelerated warming in the tropical West Pacific and cooling in the tropical and subtropical East Pacific, that has led to a progressively more negative feedback \cite{zhou_impact_2016, gregory_variation_2016, andrews_effect_2022}. This feedback change has been dominated by increases in EIS and LCC in the Eastern Equatorial Pacific and in the North- and South-Eastern subtropical Pacific low cloud decks \cite{zhou_impact_2016, ceppi_relationship_2017, silvers_diversity_2018, andrews_effect_2022, myers_observational_2023, tam_meteorological_2026}. 

The current paradigm for what controls EIS and LCC in the East Pacific is predicated on what is known in the atmospheric dynamics literature as the Convective Quasi-Equilibrium Weak Temperature Gradient approximation \cite<QE-WTG, e.g.,>[]{wallace_effect_1992,sobel_enso_2002,zhang_how_2020,williams_circus_2023}: the free troposphere over the warm pool is in quasi equilibrium with the SSTs in the deeply convecting regions following a moist adiabatic lapse rate \cite{raymond_regulation_1995}, while the rest of the tropical free troposphere is connected to the free troposphere over the warm pool by a weak temperature gradient \cite{sobel_weak_2001}. This QE-WTG-based paradigm for the pattern effect is summarized in Chapter 7, section 7.4.4.3 of the sixth assessment report of the Intergovernmental Panel on Climate Change \cite<IPCC AR6,>[]{forster_earths_2023}:
\begin{displayquote}
SSTs in regions of deep convective ascent (e.g., in the western Pacific warm pool) govern the temperature of the tropical free troposphere and, in turn, affect low-clouds through the strength of the inversion that caps the boundary layer (i.e., the lower-tropospheric stability) in subsidence regions \cite{wood_relationship_2006,klein_low-cloud_2017}. Surface warming within ascent regions thus warms the free troposphere and increases LCC, causing an increase in emission of thermal radiation to space and a reduction in absorbed solar radiation.
\end{displayquote}
QE-WTG thus presents us with a simple rule: warming in ascent regions leads to increased stability, increased LCC, and negative radiation anomalies in the cloud-capped tropical and subtropical East Pacific boundary layers. This rule is invoked in many papers \cite<e.g.,>[ and others]{zhou_impact_2016,mauritsen_clouds_2016,ceppi_relationship_2017,fueglistaler_observational_2019,rugenstein_patterns_2023,fueglistaler_peculiar_2021,alessi_surface_2023,williams_circus_2023} usually accompanied by variations on Fig. 7.14 in IPCC AR6.

\begin{figure}
\noindent\includegraphics[width=\textwidth]{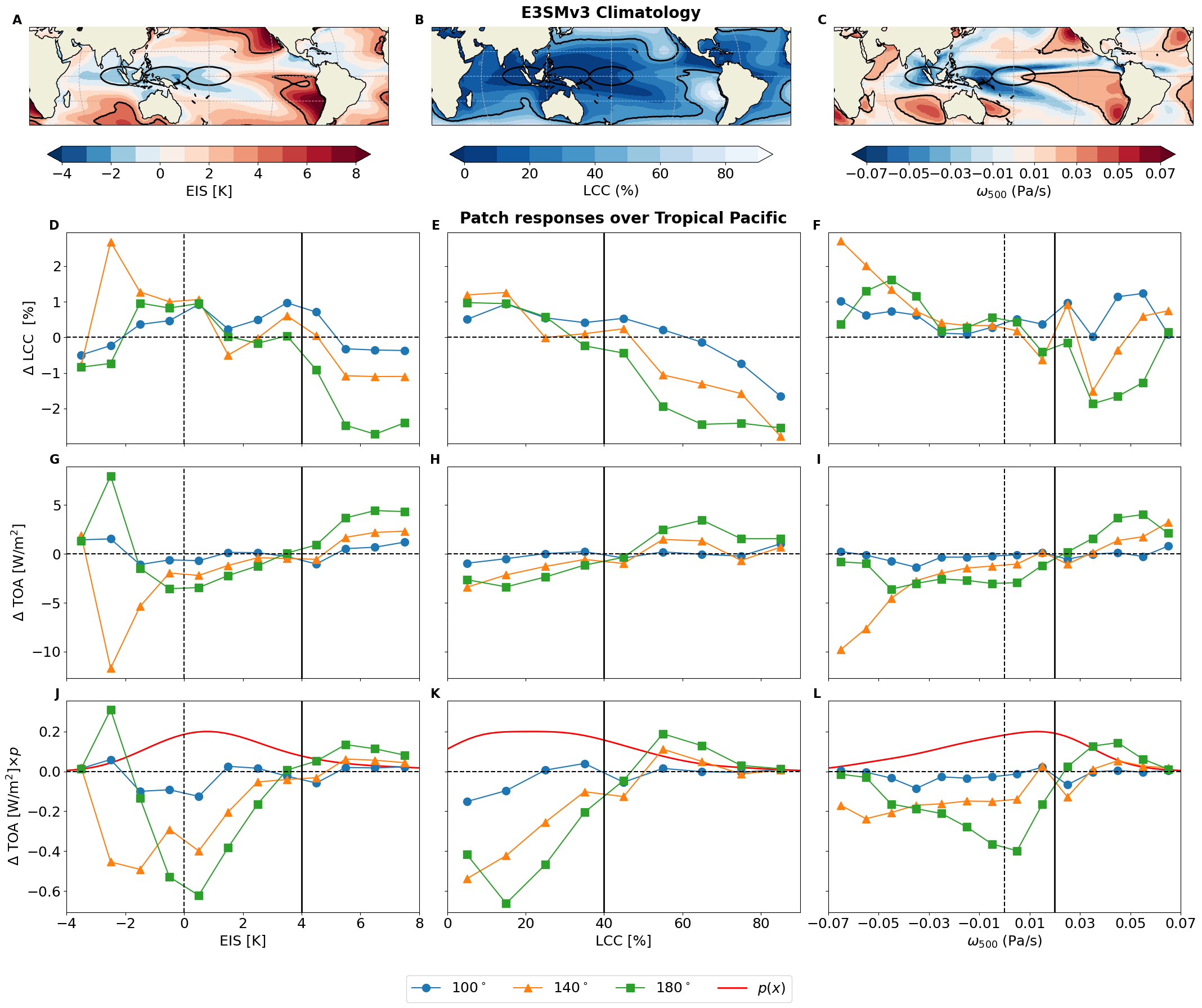}
\caption{\textbf{Bony Decomposition}: \textbf{A-C}: Climatology of EIS, LCC, and $\omega_{500}$ in the E3SMv3 control simulation. Ellipses indicate equatorial warming patches applied in three Green's Function simulations. \textbf{D-F}: Tropical Pacific response of LCC to three patches, centered at 100$^\circ$W (blue circles), 140$^\circ$W (orange triangles), and 180$^\circ$W (green squares). The responses are time- and spatially-averaged the Pacific basin between 35$^\circ$S and 35$^\circ$N, and binned by the climatological values of EIS, LCC, and $\omega_{500}$. Bin edges for the Bony plots (D-L) are identical to the contour edges in climatology plots (A-C). Black vertical lines in Bony plots correspond to the black contour in the climatology. \textbf{G-I}: Same as D-F, but for the tropical response of net TOA radiation to the three patches. \textbf{J-L}: Same as D-F, but where TOA is weighted by the frequency of each EIS, LCC, and $\omega_{500}$ bin (red line, computed using Kernel Density estimation). The area under each curve is thus equal to the average change over the Tropical Pacific.} 
\label{fig:bony}
\end{figure}

\section{The insufficiency of QE-WTG}

We highlight the insufficiency of the QE-WTG paradigm in explaining the low cloud response with analysis of the Energy Exascale Earth System Model version 3 (E3SMv3) Green's Function \cite{wu_applying_2025}. We focus on three warming patches centered on the equator and $100^\circ$E, $140^\circ$E, and $180^\circ$E, and look at the changes in LCC and TOA radiation over the tropical Pacific between 35S and 35N. We perform a Bony decomposition \cite{bony_marine_2005}, binning these responses according to climatological values of EIS, LCC, and vertical velocity at 500 hPa, $\omega_{500}$, (Figure \ref{fig:bony}A-C).

Contrary to QE-WTG predictions, all three equatorial warming patches lead to a decrease in LCC and positive radiation anomalies in regions of strongly capped boundary layers (Fig. \ref{fig:bony}D,G). Binning by climatological values of LCC similarly shows a decrease in LCC and associated positive radiation anomalies in the East Pacific subtropical cloud decks(Fig. \ref{fig:bony}E,H), while binning by circulation regimes shows positive radiation anomalies in regions of moderate to strong subsidence (Fig. \ref{fig:bony}I), although the sign of $\Delta$LCC in these regions varies by patch location and speed of descent (Fig. \ref{fig:bony}F), likely because the subsidence regions extend beyond the low cloud decks.

So what then explains the consistent result that warm pool warming leads to an overall negative radiation response when averaged over the Tropical Pacific?  Although the low cloud decks give a positive radiation response, they occupy a small fraction of the tropics. Instead, the overall negative radiation response is dominated by regions with near zero climatological EIS, low climatological LCC, and characterized by ascent or weak descent, which occupy a significantly larger area (Fig. \ref{fig:bony}J-L). 

\section{The forgotten role of wave dynamics}

Although QE-WTG has been used as a simple unifying explanation in the pattern effect literature, it's insufficiency is unsurprising, and tropical wave dynamics suggests things are indeed more complicated.  QE-WTG is expected to hold primarily between 10S and 10N \cite{sobel_weak_2001}. The low cloud decks lie outside of this region, where the Coriolis parameter is large enough to support wave-mediated horizontal gradients \cite{sardeshmukh_generation_1988, bao_zonal_2022,adames_basic_2022}. Rossby waves are also known to be important in setting up the climatology of the low cloud decks \cite{takahashi_processes_2007,wallace_atmospheric_2023}

Greens Function simulations confirm the importance of wave dynamics. Warm pool warming leads to a decrease in stability over parts of the East Pacific cloud decks in every Green's Function simulation where stability is plotted:  HadGEM2-A \cite[Fig. 5a]{andrews_dependence_2018}, CAM5 \cite[Fig. 2k]{zhou_analyzing_2017}, CAM4 \cite[Fig. 4a]{dong_attributing_2019}, and ICON \cite[Fig. S5]{williams_circus_2023}. The EIS pattern in all these simulations is reminiscent of the classical stationary Rossby wave responses to localized heating \cite{hoskins_steady_1981}, which has been previously shown to lead to significant responses in the subtropical East Pacific \cite{sardeshmukh_generation_1988}. Interestingly, two of the first papers analyzing remote radiative responses to localized West Pacific warming \cite{andrews_dependence_2018,zhou_analyzing_2017} both mention wave trains in their interpretation of the EIS response. 

\begin{figure}
\noindent\includegraphics[width=\textwidth]{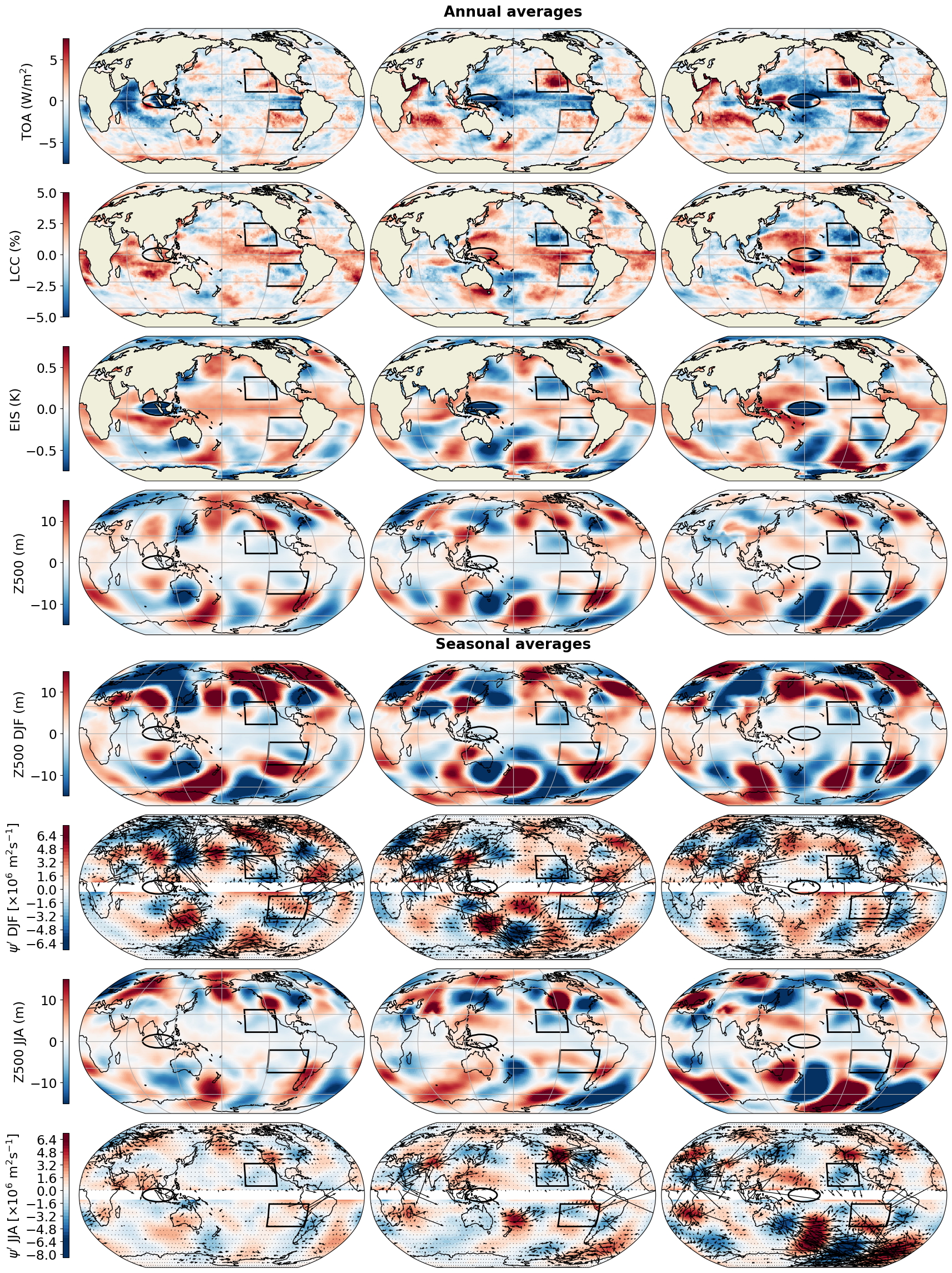}
\caption{\textbf{Greens Function Responses:} Responses of TOA radiation, LCC, EIS, Z500, and quasi-geostrophic stream function anomaly $\psi'$ \cite{plumb_three-dimensional_1985} to three equatorial warming patches centered at 100$^\circ$W (left column), 140$^\circ$W (middle column), and 180$^\circ$W (right column). The corresponding \citeA{plumb_three-dimensional_1985} wave activity flux is indicated by the vector arrows. The locations of the warming patches are indicated by the black ellipses. Top plots show annual average (ANN) anomalies  for TOA, LCC, and EIS, and Z500. Due to seasonal differences in wave propagation, lower plots show boreal winter (DJF) and boreal summer (JJA) anomalies  for Z500, $\psi'$ and wave activity flux.}
\label{fig:rossbywaves}
\end{figure}

We supplement these past results with additional analysis of the E3SMv3 Green's Function in Figure \ref{fig:rossbywaves}. Consistent with the Bony plots, but contrary to QE-WTG predictions, we find positive TOA anomalies, and decreases in LCC and EIS over the low cloud decks in response to all three patches. The decreases in EIS correspond to negative anomalies in 500 hPa geopotential height (Z500), and both of them have characteristic wave train patterns. A quasi-geostrophic stationary wave flux analysis following \citeA{plumb_three-dimensional_1985} supports the dominant role of Rossby waves in setting the pattern of Z500 and EIS in the low cloud decks. 

The QE-WTG prediction of increased EIS does hold well in the deep tropics. A narrow band of increased LCC and EIS can be observed everywhere in the Pacific between roughly 10$^\circ$S to 10$^\circ$N, and extending even further in the central and West Pacific. Despite it's varying latitudinal extent, it is this broad increase in EIS and LCC in the near-equatorial region that explains the negative tropically-averaged radiative anomalies, rather than the response in the relatively smaller low cloud decks.

\section{Discussion}
The QE-WTG paradigm presents a tantalizingly simple explanation for the pattern effect: warm pool warming results in increased EIS and LCC over the Tropical Pacific. However, Greens Function simulations suggest that localized warm pool warming actuates large scale waves that results in decreased EIS and LCC over the East Pacific low cloud decks. Interestingly, the overall prediction of the QE-WTG paradigm still holds, due to the consistent impact on EIS in the deep tropics. 

Wave dynamics preclude a simple picture for the pattern effect. The wave response to more realistic SST patterns cannot be easily deduced from localized warming experiments, so it remains unclear what should be the contribution of recent warm pool warming to recent changes in the East Pacific low cloud decks. Still, these result call into question interpretations of the low cloud deck increase over recent decades, which has at least partially been attributed to amplified warm pool warming \cite{zhou_impact_2016, mauritsen_clouds_2016}. Interestingly the two studies that tried to directly attribute changes in low clouds to specific regional warming found correlations with local cooling, but not with warm pool warming \cite{mackie_contrasting_2021, kang_recent_2023}. 

The phase of the wave response over the low cloud decks is relatively insensitive to the location of the equatorial heating, consistent with past work by \citeA{sardeshmukh_generation_1988}. Still,  stationary wave patterns are sensitive to the structure of the background flow \cite<e.g.,>[]{hoskins_rossby_1993,jin_direct_1995}. Model biases in this background flow can thus result in biases in the resulting teleconnections patterns \cite{raymond_regulation_1995,li_effect_2020}, and therefore biases in the response of the low cloud decks. Additionally, while both Z500 and the wave flux analysis suggest a strong role for large-scale stationary waves, the precise nature of these waves is not clear from our simple analysis. For example, the areas of strongest response (rectangles in Figure \ref{fig:rossbywaves}) seem to be part of a secondary wave train, at least for certain seasons and patch locations. It is possible that the low cloud decks are sensitive to both refracting extratropical Rossby waves, tropical Rossby and Kelvin waves, as well as cross-equatorial propagating waves. 

Future work will be needed to better understand how exactly these critical low cloud decks respond to remote warming, especially for more realistic warming patterns. Ultimately, simple heuristics such as can be derived from QE-WTG may be insufficient, and wave models will be needed to understand the impact of wave dynamics on the low cloud decks. A rich hierarchy of such models exists, both linear and nonlinear, moist and dry, barotropic and baroclinic \cite<e.g.,>[]{hoskins_rossby_1993,ting_steady_1998,li_interhemispheric_2015,watanabe_moist_2003}, which would ideally be paired with the Green's Function Model Intercomparison Project \cite<GFMIP,>[]{blochjohnson_greens_2024}.

\section*{Open Research Section}
All data and code to reproduce the results shown is available at 

https://doi.org/10.5281/zenodo.19211964

\section*{Conflict of Interest declaration}
The authors declare there are no conflicts of interest for this manuscript.

\acknowledgments
CP and PP were supported by the Department of Energy (DOE) Award \# DE-SC0022110 through the Regional and Global Model Analysis (RGMA) program. CP was supported by the NASA New Investigator Program grant \# 80NSSC21K1043. MFS was supported by DOE Award \# DE-SC0025595. AILW was supported by the Scripps Institutional Postdoctoral Fellowship. This is IPRC publication X and SOEST contribution Y. We thank Shiv Priyam Raghuraman and David Battisti for feedback. Claude Sonnet 4.6 was used to claw the desired layout out of matplotlib.
% MFS: X and Y are placeholders. The numbers are issued by my institution upon acceptance of the paper and can be updated during the proof stage typically.

%%%%%%%%%%%%%%%%%%%%%%%%%%%%%%%%%%%%%%%%%%%%%%%
% REFERENCES and BIBLIOGRAPHY
%
% \bibliography{<name of your .bib file>} don't specify the file extension
% don't specify bibliographystyle
%
%%%%%%%%%%%%%%%%%%%%%%%%%%%%%%%%%%%%%%%%%%%%%%%

\bibliography{RossbyWaves_PatternEffect}

%Reference citation instructions and examples:
%
% Please use ONLY \cite and \citeA for reference citations.
% \cite for parenthetical references
% ...as shown in recent studies (Simpson et al., 2019)
% \citeA for in-text citations
% ...Simpson et al. (2019) have shown...
%
%
%...as shown by \citeA{jskilby}.
%...as shown by \citeA{lewin76}, \citeA{carson86}, \citeA{bartoldy02}, and \citeA{rinaldi03}.
%...has been shown \cite{jskilbye}.
%...has been shown \cite{lewin76,carson86,bartoldy02,rinaldi03}.
%... \cite <i.e.>[]{lewin76,carson86,bartoldy02,rinaldi03}.
%...has been shown by \cite <e.g.,>[and others]{lewin76}.
%
% apacite uses < > for prenotes and [ ] for postnotes
% DO NOT use other cite commands (e.g., \citet, \citep, \citeyear, \nocite, \citealp, etc.).
%

\end{document}